\documentstyle[psfig]{mn}

\newif\ifAMStwofonts
\AMStwofontstrue

\ifoldfss
  \newcommand{\rmn}[1] {{\rm #1}}

  \ifCUPmtlplainloaded \else
    \NewTextAlphabet{textbfit} {cmbxti10} {}
    \NewTextAlphabet{textbfss} {cmssbx10} {}
    \NewMathAlphabet{mathbfit} {cmbxti10} {} 
    \NewMathAlphabet{mathbfss} {cmssbx10} {} 
  \fi
  \ifAMStwofonts
    \ifCUPmtlplainloaded \else
      \NewSymbolFont{upmath} {eurm10}
      \NewSymbolFont{AMSa} {msam10}
      \NewMathSymbol{\upi}     {0}{upmath}{19}
      \NewMathSymbol{\umu}     {0}{upmath}{16}
      \NewMathSymbol{\upartial}{0}{upmath}{40}
      \NewMathSymbol{\leqslant}{3}{AMSa}{36}
      \NewMathSymbol{\geqslant}{3}{AMSa}{3E}

      \let\leq=\leqslant 
      \let\geq=\geqslant 
    \fi
  \fi
\fi 

\ifnfssone
  \newmathalphabet{\mathit}
  \addtoversion{normal}{\mathit}{cmr}{m}{it}
  \addtoversion{bold}{\mathit}{cmr}{bx}{it}
  \newcommand{\rmn}[1] {\mathrm{#1}}

  \newmathalphabet{\mathbfit} 
  \addtoversion{normal}{\mathbfit}{cmr}{bx}{it}
  \addtoversion{bold}{\mathbfit}{cmr}{bx}{it}
  \newmathalphabet{\mathbfss} 
  \addtoversion{normal}{\mathbfss}{cmss}{bx}{n}
  \addtoversion{bold}{\mathbfss}{cmss}{bx}{n}
  \ifAMStwofonts
    \ifCUPmtlplainloaded \else
      \UseAMStwoboldmath
      \makeatletter
      \new@mathgroup\upmath@group
      \define@mathgroup\mv@normal\upmath@group{eur}{m}{n}
      \define@mathgroup\mv@bold\upmath@group{eur}{b}{n}
      \edef\UPM{\hexnumber\upmath@group}
      \new@mathgroup\amsa@group
      \define@mathgroup\mv@normal\amsa@group{msa}{m}{n}
      \define@mathgroup\mv@bold\amsa@group{msa}{m}{n}
      \edef\AMSa{\hexnumber\amsa@group}
      \makeatother
      \mathchardef\upi="0\UPM19
      \mathchardef\umu="0\UPM16
      \mathchardef\upartial="0\UPM40
      \mathchardef\leqslant="3\AMSa36
      \mathchardef\geqslant="3\AMSa3E

      \let\leq=\leqslant 
      \let\geq=\geqslant 
    \fi
  \fi
\fi 

\ifnfsstwo
  \newcommand{\rmn}[1] {\mathrm{#1}}

  \DeclareMathAlphabet{\mathbfit}{OT1}{cmr}{bx}{it}
  \SetMathAlphabet\mathbfit{bold}{OT1}{cmr}{bx}{it}
  \DeclareMathAlphabet{\mathbfss}{OT1}{cmss}{bx}{n}
  \SetMathAlphabet\mathbfss{bold}{OT1}{cmss}{bx}{n}
  \ifAMStwofonts
    \ifCUPmtlplainloaded \else
      \DeclareSymbolFont{UPM}{U}{eur}{m}{n}
      \SetSymbolFont{UPM}{bold}{U}{eur}{b}{n}
      \DeclareSymbolFont{AMSa}{U}{msa}{m}{n}
      \DeclareMathSymbol{\upi}{0}{UPM}{"19}
      \DeclareMathSymbol{\umu}{0}{UPM}{"16}
      \DeclareMathSymbol{\upartial}{0}{UPM}{"40}
      \DeclareMathSymbol{\leqslant}{3}{AMSa}{"36}
      \DeclareMathSymbol{\geqslant}{3}{AMSa}{"3E}

      \let\leq=\leqslant 
      \let\geq=\geqslant 
    \fi
  \fi
\fi 

\ifCUPmtlplainloaded \else
  \ifAMStwofonts \else 
    \def\upi{\pi}
    \def\umu{\mu}
    \def\upartial{\partial}
  \fi
\fi

\title{Evolutionary Constraints from Infra-Red Source Counts}
\author[Chris P. Pearson]
       {Chris P. Pearson$^1$\thanks{Further information contact Chris Pearson (cpp@ir.isas.ac.jp)         $http://www.ir.isas.ac.jp/\sim cpp/counts/$ }\\
        $^1$Institute of Space and Astronautical Science, Yoshinodai 3-1-1, Sagamihara, Kanagawa 229 8510, Japan}

\date{Accepted .\\
      Received ;\\
      in original form 2000 November 10}
\pagerange{\pageref{firstpage}--\pageref{lastpage}}

\begin{document}

\label{firstpage}

\maketitle

\begin{abstract}

The {\it backward evolution} approach to modelling galaxy source counts is re-visited in the wake of the numerous results and revelations from the Infrared Space Observatory (ISO), the Submillimetre Common User Bolometer Array (SCUBA) and the detections and measurements of the cosmic extragalactic background light. Using the framework of the Pearson \& Rowan-Robinson (1996) galaxy evolution model, the observed source counts and background measurements are used to constrain the evolution in the galaxy population. It is found that a strong evolution in both density and luminosity of the high luminosity tail of the IR luminosity function, interpreted as the ultraluminous galaxies discovered first by IRAS and later elevated in status by SCUBA and ISO, can account for the source counts from 15$\umu$m (where it matches the undulations in the integral counts and hump in the differential counts extremely well), to the sub-mm region, as well as explain the peak in the cosmic infrared background at $\sim$140$\umu$m. The sub-mm counts are interpreted as the superposition of 2 separate populations comprising of ULIGs at the brighter sub-mm fluxes and starburst galaxies at fluxes fainter than $\sim$2mJy. In this scenario the high redshift ULIGs are tenuously interpreted as the progenitors of today's giant elliptical galaxies.

All the source count models can be accessed via the world wide web at this URL :

$http://www.ir.isas.ac.jp/\sim cpp/counts/$

\end{abstract}


\section{Introduction}\label{sec:introduction}

The formation epoch of galaxies has long been the {\it holy grail } of cosmology. Traditionally one of the most popular and successful ways of glimpsing this epoch has been the analysis of extragalactic source counts or more commonly galaxy number counts. Although originally intended as a means of determining the geometry of the Universe this tool has shown to be much more viable and useful in the study of the evolutionary and star formation history of galaxies (Kirshner et al.\shortcite{kirshner81}, Ellis\shortcite{ellis87}). In particular, the large dust content of galaxies discovered by IRAS and the strong K-corrections have made the infra-red and more recently, the sub-mm, extremely important wavebands for the study of galaxy evolution and cosmic star formation. By modelling the counts of galaxies it has become possible to analyze the evolution of the galaxy population out to redshifts of unity and higher. Coupled with the advent of new observational data, galaxy number counts are poised to discriminate between rivalling evolutionary theories.

Recently, it has become the vogue to tackle the problem of the number distribution of galaxies in the Universe via one of two methods. The first (often called the {\it backward evolution} approach \cite{lonsdale96}), takes the observed, present day ($z=0$) luminosity function and evolves it in luminosity and/or density back out to higher redshifts assuming some parameterization of the evolution (Beichman \& Helou \shortcite{beich91}, Blain \& Longair \shortcite{blain93}, Pearson \& Rowan-Robinson \shortcite{cpp96}, Xu et al. \shortcite{xu98}, Takeuchi et al. \shortcite{take99}). This method has the advantages that it is both direct and relatively simple to implement with few free parameters and assumptions about the Universe at earlier times. The disadvantage in the past was that the information on which the assumptions about evolution were made often came from IRAS data which extended out only to relatively low redshifts $\approx 0.02-0.2$. In the alternative method, known as the {\it forward evolution} approach the evolution is computed assuming some initial conditions of the physical processes of chemical evolution and photometric evolution of the stellar populations that heat the dust \cite{fran94}. The so-called {\it semi analytical approach} combines these assumptions on the chemical/photometric evolution of galaxies with models considering the dissipative and non-dissipative processes affecting galaxy formation within dark matter haloes (Cole et al. \shortcite{cole94}, Baugh et al \shortcite{baugh96}) and has provided a reasonable fit to the source counts in the infra-red \cite{guid98}. The {\it forward evolution} approach has the advantage of being based on perhaps a more fundamental set of assumptions, however the obvious disadvantages are the larger number of free/unknown parameters assumed in these models. 

Furthermore, with the recent advent of large, deep surveys by SCUBA (Holland et al. \shortcite{holland99}, e.g.  Smail et al.\shortcite{smail97}, Hughes et al.\shortcite{hugh98}, Barger et al.\shortcite{barg98}, Blain et al.\shortcite{blain99}, Barger et al.\shortcite{barg99}) and ISO (Kessler et al.\shortcite{kessler96}, e.g. Bogun et al.~\shortcite{bogun96}, Serjeant et al.~\shortcite{serjeant97}, Taniguchi et al.~\shortcite{taniguchi97}, Elbaz et al.~\shortcite{elbaz98}, Stickel et al.~\shortcite{stick98}, Kawara et al.~\shortcite{kawara98}, Flores et al.~\shortcite{flores99a},~\shortcite{flores99b}, Altieri et al.~\shortcite{altieri99}, Oliver et al.~\shortcite{oliver00a}, Lindern-V{\o}rnle et al.~\shortcite{vornle00}, Oliver et al.~\shortcite{oliver00b}), including constraints on the source counts to fainter levels and the IR background (Lagache et al.\shortcite{lagache99},\shortcite{lagache00}, Matsuhara et al. \shortcite{mat00}, Matsumoto et al.\shortcite{mats00}) and detections by the COBE FIRAS/DIRBE instruments at 140 \& 240$\umu$m (Puget et al. \shortcite{pug96}, Fixsen et al. \shortcite{fixsen98}, Hauser et al. \shortcite{hauser98}) the tools are now available to constrain the {\it backward evolution} methodology to significantly higher redshifts. In the light of these recent advances in the mm-IR region the {\it backward evolution} method is re-visited.

The aim of this paper is not to produce a definitive evolutionary model to fit all observed counts while satisfying various preferences on the various dominant populations (starburst, AGN, luminous infrared galaxies (LIG's $L_{IR} > 10^{11}L_{\sun}$), ultraluminous galaxies (ULIG's $L_{IR} > 10^{12}L_{\sun}$), hyperluminous galaxies ($L_{IR} > 10^{13}L_{\sun}$)). Rather I seek to investigate the various avenues of evolution that can viably fit the galaxy source counts from the mid-IR to sub-mm wavelengths whilst not violating the constraints set by the IR background and CMB measurements. The paper is structured as follows, in section \ref{sec:model} the galaxy count and world model parameters are outlined. Section \ref{sec:evol} reviews current evolutionary models and constraints and analyszes simple density and luminosity evolution scenarios. Section \ref{sec:counts} formulates a new evolutionary source count model and applies it to the observed source counts and background measurements from the mid-IR to sub-mm wavelengths. The possible interpretations and conclusions are reported in sections \ref{sec:discuss} \& \ref{sec:conc} respectively.

\section{Model Parameters}\label{sec:model}

In general, for observations at frequency $\nu = c/ \lambda $, the total number of sources per steradian observable down to a flux sensitivity $S$ is given by;

\begin{equation}
   N(S_{\nu})=\!\!\int_{0}^{\infty}\!\! \ \!\!\int_{0}^{z(L,S)}\!\! \phi (L/f(z))  {dV(z)\over{dz}}g(z) {\rmn{d}}lgL {\rmn{d}}z,
\end{equation}

\begin{equation}
   {dV(z)\over{dz}} = {4\upi D^2 \over{ H_{o}(1+z)^2(1+\Omega_{o} z)^{1/2}}},
\end{equation}

where $\phi = d \Phi / dlgL = $ the zero redshift differential luminosity function per decade in luminosity $L$, parameterizing the number density of extragalactic sources as a function of luminosity at a frequency $\nu$. ${\rmn{d}}V/{\rmn{d}}z$ is the general form of the differential volume element with redshift, $z$, and is dependent on the luminosity distance, $D$, and assumed cosmological world model. $f(z)$ \& $g(z)$ are luminosity and density evolutionary parameters respectively. The limiting redshift $z(L,S)$, is determined by the flux, $S$, below which a source of luminosity, $L$, becomes too weak to be included in the sample of galaxies observed, where;

\begin{equation}
   S(\nu_{o})= {L_{\nu_{o}} \over{4\pi D^{2}}} {\nu_{e} L_{\nu_{e}} \over{\nu_{o} L_{\nu_{o}}}} f(z) ,
\end{equation}

\begin{equation}
   D(z) = {2c\over{ H_{o}\Omega_{o}^{2} }} (\Omega_{o} z+(\Omega_{o} -2)[(1+\Omega_{o} z)^{1/2}-1]),
\end{equation}

where ${\nu_{e} L_{\nu_{e}} \over{\nu_{o} L_{\nu_{o}}}}$ is effectively the ratio of the emission luminosity (luminosity at $\nu_{e}$) to the observed luminosity (i.e. the K-correction).
Even assuming a single luminosity function and evolutionary history for all extragalactic sources can in fact come impressively close to matching the true numbers and distribution of sources over a wide range of wavelengths as observed in the Universe (e.g. Takeuchi et al. \shortcite{take99}, ~\shortcite{take00}).

Pearson \& Rowan-Robinson~\shortcite{cpp96} (hereafter cppRR) provided an impressively solid framework from which was modelled the source counts from radio frequencies, sub-mm, IR, optical, UV wavelengths to X-ray energies. The original IR models of cppRR consisted of a  4 component parameterization of the extragalactic population based on the IR colours of IRAS galaxies ~\cite{RR89}. The model incorporated a normal galaxy (IR cirrus dominated) component defined by cool 100/60$\umu$m colours, a starburst component defined by warm 100/60$\umu$m colours extending to an ultra/hyperluminous component at higher IR luminosities and an AGN Seyfert/QSO 3-30$\umu$m dust torus component defined by low 60/25$\umu$m colours. The basic framework of this model has been maintained in this further analysis.

The original template spectral energy distributions (SED's) of Rowan-Robinson \shortcite{RR92}, Rowan-Robinson \& Efstathiou \shortcite{RR931} for the {\it cool} normal, {\it warm} starburst \& ultraluminous galaxies are replaced by the new radiative transfer models of  Efstathiou \& Siebenmorgen~\shortcite{esf002} and Efstathiou, Rowan-Robinson \& Siebenmorgen~\shortcite{esf001} respectively. The new starburst galaxy models are calculated considering the evolution of an ensemble of optically thick giant molecular clouds, illuminated by embedded massive stars. The starburst models are defined by the parameters {\it t\ }, the age of the starburst in Myrs; $\tau$, the initial optical depth (in {\it V\ }) of the molecular clouds; $\chi$, the ratio of radiation field to the local solar neighborhood. The starburst SED can be considered akin to M82, while the ULIG SED represents an ARP220 type galaxy. The introduction of this ARP220 type SED replaces the more extreme hyperluminous galaxy IRAS F10214 (Rowan-Robinson et al.~\shortcite{RR91a},~\shortcite{RR932}) that was used in the earlier models of cppRR. F10214 does not have a typical ULIG SED and such a hyperluminous galaxy population may represent an even more extreme extension of the starburst phenomenon than the ULIGs, only 39 being so far identified \cite{RR00}. Efstathiou, Rowan-Robinson \& Siebenmorgen ~\shortcite{esf001} demonstrated that exponentially decaying $10^{7} \sim 10^{8}$yrs old starbursts are capable of reproducing both the IRAS colours of galaxies and also provide good fits to the data for the stereotypical starburst galaxy M82 and to the recently ISO observed starburst galaxy NGC6090 ~\cite{acost96}.

\begin{figure}
\centering
\centerline{
\psfig{figure=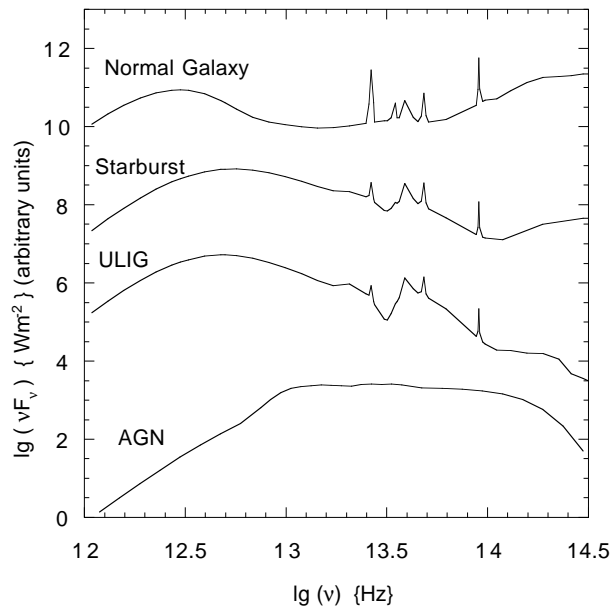,height=8cm}
}
\caption{ Model spectral energy distributions used for the source count models.
\label{sed}}
\end{figure}

 Of particular importance to the models in the mid-infrared region is the inclusion of the unidentified infra-red bands (UIB), popularly supposed to be band emission from Polycyclic Aromatic Hydrocarbons (PAH's ~\cite{pug89}), into the model SED's. These features are prominent from 3-12$\umu$m in most galaxies (~\cite{lu97}, ~\cite{boul96},~\cite{vig96}) having equivalent widths of as much as 10$\umu$m, comparable to those of the ISOCAM MIR filter band passes (3.5$\umu$m \& 6$\umu$m for the LW2(5-8.5$\umu$m) filter and LW3(12-18$\umu$m) filters respectively). These UIB's are expected to significantly distort the mid-IR source counts compared to those constructed purely on the basis of the IRAS SED data (Xu et al.~\shortcite{xu98}, Pearson et al.~\shortcite{cpp01}). A further significant and necessary improvement over the original cppRR model is the introduction of the full range of PAH features ~\cite{sieb92} that were not considered in the earlier model predictions. Fig.~\ref{sed} shows the new model SED's. The shape and form of the PAH features assumed in different models can vary significantly (e.g. see Xu et al.~\shortcite{xu98}, Dole et al.\shortcite{dole00}) but although the PAH features may seem overly strong or sharply defined, as long as the energy contained within the feature peaks is consistent with observation and the SEDs are correctly smoothed and convolved with the filter response function, the general profile of the individual features is not critical as galaxies will naturally have a range of feature profiles in their SEDs.

Note that recently Dunne et al. ~\shortcite{dunne00a} have shown from SCUBA data on local galaxies that there is a correlation between the far-IR-sub-mm SED and the 60$\umu$m luminosity function. This effect is not treated in the model SEDs and will be the subject of a later investigation.

For the AGN SED the original SED of Rowan-Robinson ~\shortcite{RR95} is retained. The MIR SED of AGN is relatively featureless compared to that of starburst galaxies, exhibiting none of the broadband UIB features present in the latter's spectrum \cite{lutz97}. The AGN central engine (black hole) would be expected to produce an extreme environment destroying or changing the smaller dust grains and their composition.

The model components are distinguished on a basis of their SED's and number-luminosity distributions (i.e. luminosity function). The {\it cool } normal galaxies, {\it warm } starbursts \& ULIG galaxies are well described locally by the IRAS 60$\umu$m luminosity function \cite{saun90} which is a log Gaussian function producing a slightly broader distribution than a traditional Schecter function and preferable to the slightly less physical power law parameterizations of IRAS galaxies \cite{soif87}. In this parameterization the normal and starburst galaxies extend in 60$\umu$m luminosity from $lgL_{60\mu m} = 8-11.5L_{\sun} $ while the ULIG population forms the high end tail of the starburst luminosity function from $lgL_{60\mu m} = 11.5-14L_{\sun} $.  The AGN are represented by the 12$\umu$m luminosity function of Lawrence et al.~\shortcite{law86} defined from $lgL_{12\mu m} = 8-14L_{\sun} $ with parameters corresponding to the 12$\umu$m sample of Rush et al.~\shortcite{rush93} for Seyfert 1 and Seyfert 2 galaxies. The parameters for the different components given in table~\ref{lf} correspond to those in the equations below for the 60$\umu$m galaxy and 12$\umu$m AGN luminosity functions respectively;

\begin{equation}
   \phi(L_{60}) = {{\rmn{d}}\Phi\over{{\rmn{d}}L}} = C({L\over{L*}})^{1-\alpha}exp[-{1\over{2\sigma ^2}} lg^2 (1+{L\over{L*}})],
\label{LF60}
\end{equation}

\begin{equation}
   \phi(L_{12}) = {{\rmn{d}}\Phi\over{{\rmn{d}}L}} = CL^{1-\alpha} (1+{L\over{L*\beta}})^{-\beta},
\end{equation}

To shift the luminosity function from the wavelength at which the luminosity function is defined, $\lambda _{LF} $, to the observation wavelength, $\lambda _{obs} $, the ratio $L(\lambda _{obs})/L(\lambda _{LF} )$ is obtained via the model template spectra. 

 The model assumes $\Omega=0.1$, $H_o = 50kms^{-1}Mpc^{-1}$. Alternative cosmologies are not examined in this work so as to ensure that a clear distinction between the evolution scenarios can be made. The maximum redshift assumed for the models is $z=7$ although assuming a higher redshift of 10 makes little difference to the results presented here.

\begin{table*}
\begin{minipage}{110mm}
\caption{Luminosity function parameters.}
\label{lf}
\begin{tabular}{@{}llllll}

 Component  & $\alpha$ & $\sigma$ & $\beta$ & $lg(L^*/L_{\sun})$ & $C$ \\
\hline
cool 60$\umu$m & $1.15$ & $0.463$ & $-$ & $9.62$ & $2.000\times 10^{-3}$ \\
warm 60$\umu$m & $1.27$ & $0.626$ & $-$ & $9.99$ & $3.250\times 10^{-4}$ \\
ULIG 60$\umu$m & $-$ & $0.1$ & $-$ & $11.6$ & $4.185\times 10^{-6}$ \\
Seyfert 1 12$\umu$m & $1$ & $-$ & $2.1$ & $9.552$ & $2.0969\times 10^{-4}$ \\
Seyfert 2 12$\umu$m & $1$ & $-$ & $2.5$ & $9.952$ & $8.0251\times 10^{-5}$ \\
\hline
\end{tabular}
\end{minipage}
\end{table*}

\section{Evolution}\label{sec:evol}

\subsection{Observational Evolutionary Evidence}

The original cppRR model assumed a simple power law luminosity evolution $L(z)=L(z=0)(1+z)^{3.1}$ for the starburst, ultraluminous and AGN components and was found to fit the IRAS data extremely well \cite{oliver92}. This evolution was motivated by similar evolutionary models at radio and X-ray wavelengths (Boyle et al. \shortcite{boyl88}, Benn et al. \shortcite{benn93}, Condon \shortcite{condon94}). Recently, similar evolution has also been observed in optically selected starburst galaxies \cite{lilly96} and in the submilliter emission from radio loud galaxies observed at 850$\umu$m by SCUBA ~\cite{arch00}.  Although this simple form of evolution still provides a satisfactory fit to source counts down to quite faint sensitivities (1mJy at 15$\umu$m, Serjeant et al. \shortcite{serjeant00}, $\sim$0.3Jy \& $\sim$0.1Jy at 170$\umu$m \& 90$\umu$m respectively), it cannot match the observed bumps and dips in the ISO 15$\umu$m counts below 1mJy and falls below the observed faint counts at far-IR wavelengths (e.g. 170$\umu$m FIRBACK; Puget et al. ~shortcite{pug99}, Lockman Hole; Kawara et al. ~\shortcite{kawara98}). Moreover, the extremely high counts reported by the SCUBA instrument at submillimetre wavelengths at fluxes brighter than $\sim$2mJy cannot be reproduced in any way by the cppRR luminosity evolution model, although it should be noted that the predictions are tantalizingly close to the observed counts at fluxes fainter than $S_{850\mu m}<2mJy$.
Furthermore, although the faintest source counts from the 60$\umu$m IRAS survey (VFSS, $\sim$120mJy Gregorich et al. \shortcite{gregor95}, Oliver et al. \shortcite{oliver95}, Bertin et al. \shortcite{bertin97}) could be well explained by the cppRR pure evolution model at these relatively bright fluxes the evolutionary constraints are extremely tenuous with many different scenarios being allowed ( the IRAS samples not probing to significantly large redshifts (QDOT sample $z_{median}=0.03$ \cite{RR91b}, VFSS analysis $z<0.3$). At redshifts higher than $\sim 0.25$, the evolution is not well constrained.

Recently, several authors have attempted to combat these shortcomings by incorporating stronger or more drastic evolution scenarios into the source count framework. Rowan-Robinson \shortcite{RR00} included evolution explicitly into the IR 60$\umu$m luminosity function re-producing many of the observables if a low density cosmology was assumed ($\Omega_{o}=0.3$). For higher values of the density parameter it was concluded that a new, strongly evolving population would have to be invoked (e.g. Franceschini et al.\shortcite{fran98}). Takeuchi et al. \shortcite{take00} treated a global luminosity evolution scenario in a stepwise, non-parametric form concluding that strong evolution to z$\sim$1 was necesscary to fit the ISO source counts at 15$\umu$m and 170$\umu$m along with the far-IR background. An advantage of the formulation of cppRR was that the 60$\umu$m luminosity function was treated as a superposition of galaxy populations providing the added flexibility of specific evolution with extragalactic class. However, these evolution scenarios must be reviewed if we are to hope to fit the SCUBA counts at 850$\umu$m and ISO 15$\umu$m \& 170$\umu$m faint counts.

 From a statistical analysis of the 60$\umu$m IRAS VFSS, Bertin et al. ~\shortcite{bertin97} concluded that the most luminous IRAS galaxies should be $\approx$ 5$\sim$7 times more numerous at z$\approx$0.3 than at the present epoch with IR luminosities of the order of $L_{60}\sim 2$x$10^{12}$ and optical magnitudes between $20<B_{J}<21$ with red colours ($B_{J}-R_{F} \approx 1.6$ making them potential ULIG candidates \cite{clements96}).
Kim \& Saunders \shortcite{kim98} compiled a complete flux limited sample of ULIGs from the IRAS Faint Source Catalogue (FSC) with a flux $f_{60}>1Jy$ to a maximum redshift of $z=0.268$ and mean redshift $<z>\sim 0.15$. Although locally ($z<0.1$), ULIGs have a space density approximately similar to optically selected QSOs, the analysis of the IRAS FSC hinted at much stronger evolution at higher redshifts to $z \approx 1$ thus outnumbering the QSOs by a factor of 2. Based on the premise that the ULIG population may follow a similar evolutionary track as QSO's \cite{hassinger98}, Kim \& Saunders \shortcite{kim98} modelled the ULIG population with density evolution of the form $(1+z)^{g}$, with $g \approx 7.6\pm 3.2$.  Similar strong density evolution (equivalent to $g \approx 10$ at $z\sim 1$) has also been proposed to fit the deep 15$\umu$m ISO source counts ~\cite{dole00}.
The vast majority of ULIGs appear to be in the latter stages of merging (98$\%$ in the sample of Borne et al.~\shortcite{borne99}, see also, Arribas et al. ~\shortcite{arribas00}, Borne et al.~\shortcite{borne00}, Colina et al.~\shortcite{colina00}) with double nuclei, overlapping disks and tidal features clearly present in many images \cite{sand98} \cite{mel90}, \cite{clements96}. The large amounts of dust present in the central regions of these galaxies provides an ideal environment for circumnuclear starburst and/or AGN adding to the growing evidence that ULIGs could be the progenitors of today's giant elliptical galaxies formed by merger induced collapse of the galaxy core, formation of globular clusters and galactic {\it superwinds} resulting in a dust free core and corresponding enrichment of the intergalactic medium  \cite{korm92}, \cite{surace98}. However, in the local Universe $ z< \sim 0.08 $, the space density of ULIGs is approximately 0.001 per sq.deg.  or $ \approx 1.1\times 10^{-7}Mpc^{-3} $ \cite{saun90}, a factor of as much as  $ \approx $1500 lower than early type galaxies \cite{trentham00}. Thus if ULIGs are to be present in sufficient numbers to be representative today's gE galaxy population then their space density must increase drastically with look back time (a factor of 1500 $ \sim 10^{-4}Mpc^{-3}$).

\begin{figure}
\centering
\centerline{
\psfig{figure=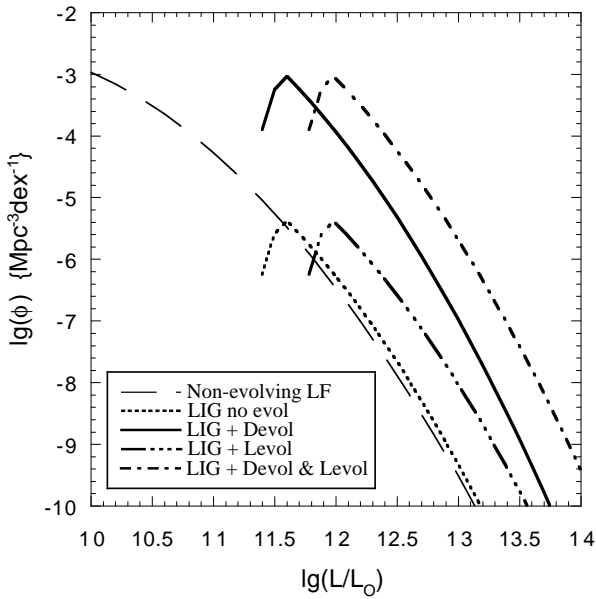,height=8cm}
}
\caption{Evolution of the high luminosity tail of the 60$\umu$m differential luminosity function. The zero redshift (no evolution) 60$\umu$m luminosity function and high luminosity tail are shown in comparison with an evolved $L^{*}=10^{11.6} L_{\sun}$ luminous infrared galaxy luminosity function tail at z=1, shown for mild pure luminosity evolution, equivalent to $(1+z)^{0.5}$, strong pure density evolution, equivalent to $\sim (1+z)^{8}$ and a combination of both. Luminosity evolution corresponds to a horizontal displacement in luminosity, density evolution corresponds to a vertical displacement in number density. 
\label{lfevol}}
\end{figure}

\subsection{Evolution of the high luminosity tail of the IR luminosity function}

\begin{figure*}
\centering
\centerline{
\psfig{figure=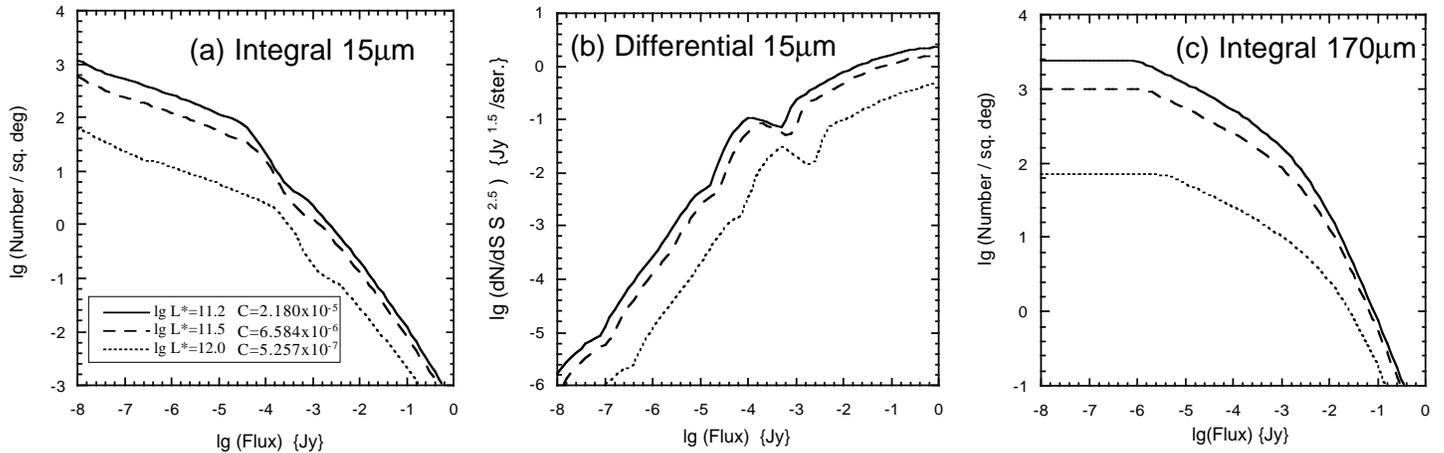,height=6cm}
}
\caption{Integral source counts of luminous/ultraluminous IR galaxies at {\it (a)} 15$\umu$m, differential counts at  {\it (b)} 15$\umu$m and integral counts at {\it (c)} 170$\umu$m, assuming values of  $lgL^{*}=11.2, 11.5, 12 L_{\sun}$. The values of {\it C} quoted are the corresponding normalizations of the luminosity function given in equation \ref{LF60} and table \ref{lf} . The shift with increasing $lgL^{*}$ to the lower right in the $lgN / lgS$ plane reflects the higher brightness but lower number density (normalization) of the sources.
\label{lstar}}
\end{figure*}

Under this motivation, extreme number evolution of the ULIG population is considered, i.e. the high luminosity tail of the 60$\umu$m luminosity function. Fig.\ref{lfevol} shows the effect of such evolution on the 60$\umu$m differential luminosity function. Evolution in the number of galaxies with redshift (density evolution) produces a vertical shift in the luminosity function plane. Evolution in the luminosity of galaxies (luminosity evolution) produces a horizontal shift in the luminosity function plane. In reality a combination of the two scenarios may well be realized. In the figure, only evolution in the high luminosity tail of the luminosity function is shown. Here I have defined a population of luminous infrared galaxies (LIG/ULIG) with $lg L^{*} \approx 11.6 L_{\sun}$, i.e. corresponding to a far-IR luminosity (50-100$\umu$m) of $ \sim 11.8 L_{\sun}$ such that the total IR luminosity (1-1000$\umu$m) of $ \sim 12 L_{\sun}$ ~\cite{helou84}.  The definition of $L^{*}$ is given in equations 5 \& 6 which define the luminosity functions used in these models. Although the luminosity functions are not classical Schecter functions, the definition of $L^{*}$ is almost identical. Changing the value of $L^{*}$ to higher luminosities shifts the luminous infra-red population down and to the right (towards a lower normalization) in the luminosity function plane and has a similar effect on the corresponding calculated source counts as shown in figure~\ref{lstar} where a brighter $L^{*}$ results in detections at brighter sensitivity limits but at a lower number density (i.e. more powerful but scarcer sources). It should be kept in mind that the main aim of this work is to try to highlight the constraints, by means of simple evolutionary scenarios, on the evolution of the high end tail of the 60$\umu$m IR luminosity function (luminosity functions in other wavebands are also avalible e.g. recently the sub-mm luminosity function has been calculated by Dunne et al.~\shortcite{dunne00a} using the SCUBA Local Universe Galaxy Survey, SLUGS).

\begin{figure*}
\centering
\centerline{
\psfig{figure=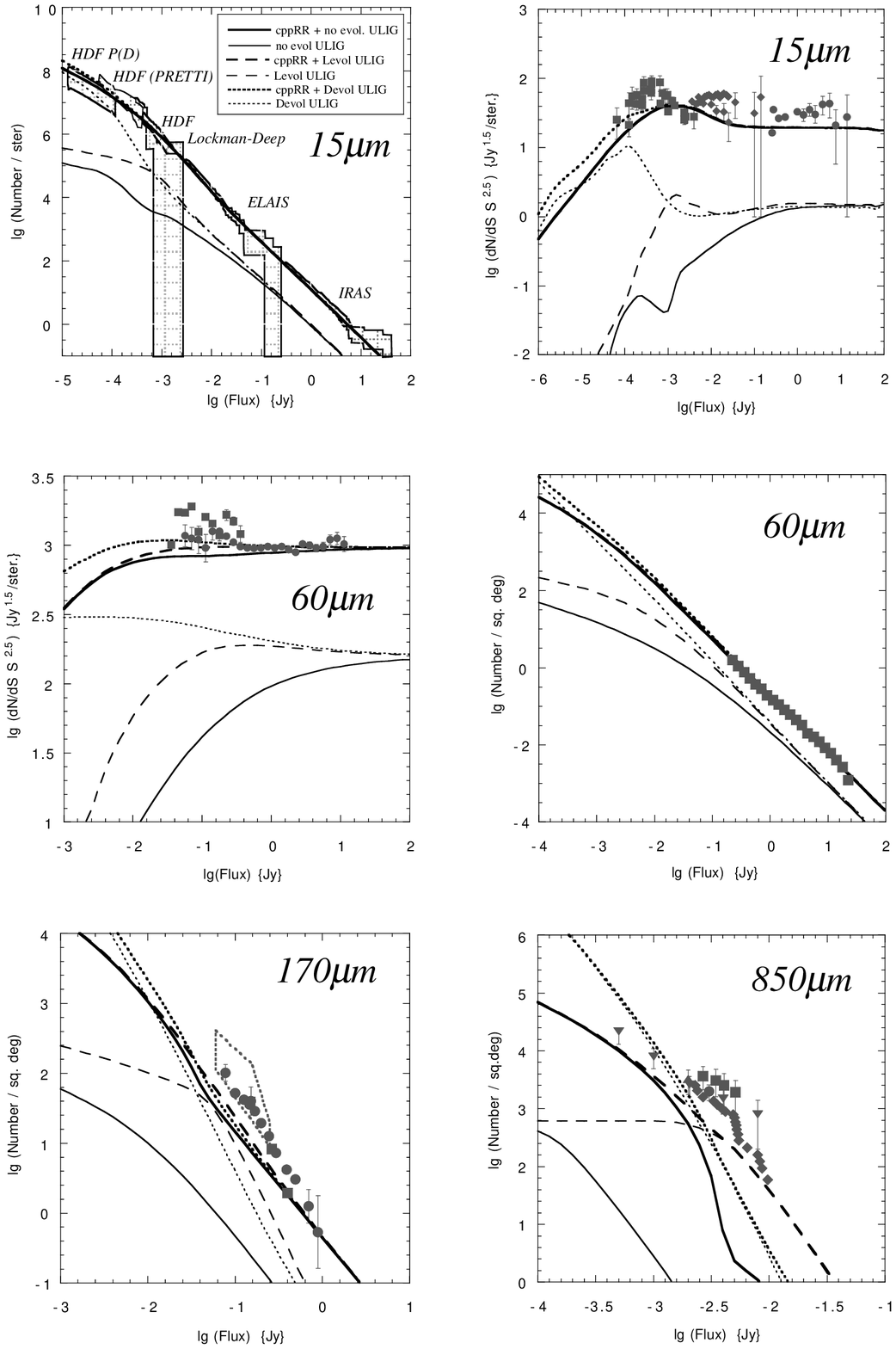,height=20cm}
}
\caption{Comparison of source counts including the luminous/ultraluminous IR galaxy (ULIG) component assuming the cppRR framework for the normal, starburst and AGN components. Integral and differential counts at 15$\umu$m, 60$\umu$m, 170$\umu$m \& 850$\umu$m are shown for 3 classic evolution scenarios of the luminous/ultraluminous IR galaxy component ({\it thin lines\,}). Also shown is the total contribution of all components including the ULIG component ({\it thick lines\,}). A non-evolving ULIG component ({\it solid lines\,}). Pure luminosity evolution of the form  $f(z)=(1+z)^{3.1}$ ({\it Levol - dashed lines\,}) and pure density evolution of the form $g(z)=(1+z)^{7}$ ({\it Devol - dotted lines\,}). Data is from 15$\umu$m - Oliver et al. (1997), Aussel et al. (1999), Serjeant et al. (2000), Elbaz et al. (1999). 60$\umu$m - Lonsdale et al. (1990), Hacking \& Houck (1987), Rowan-Robinson et al. (1990), Saunders (1990), Gregorich et al. (1995). 170$\umu$m - Kawara et al. (1998), Puget et al. (1999), Matsuhara et al. (2000) (fluctuation analysis - {\it dotted box\,}) . 850$\umu$m - Smail et al. (1997), Hughes et al. (1998), Eales et al. (1999), Barger et al. (1999), Blain et al. (1999).
\label{classics}} 
\end{figure*}  

As a starting point, I compare the pre-ISO view of galaxy evolution with the galaxy source counts observed at 15, 60, 170 \& 850$\umu$m by ISO, IRAS, ISO and SCUBA respectively. Three scenarios are presented for the evolution of the ULIG component, (a) no-evolution, (b) luminosity evolution of the form $f(z)=(1+z)^{3.1}$ (cppRR model) and (c) density evolution of the form $g(z)=(1+z)^{7}$ (e.g. Oliver et al. \shortcite{oliver95}). The evolution is continued out to a redshift of 2.5 where it is assumed to level off and eventually decline towards higher redshift. Fig.~\ref{classics} shows the individually evolving ULIG components plus the total source counts corresponding to each scenario. The total source counts include the ULIG component and the normal, starburst, AGN components (evolving as $L(z)=f(z)L(0)$ as per the cppRR model framework). From this simple representation of the evolution we can already ascertain important boundary conditions on the evolution of the ULIG component. At the brightest fluxes all models agree on the same normalization as they should seeing that at these fluxes the counts do not probe deep enough to discriminate between the different evolution scenarios. Both the luminosity and density evolution scenarios are monotonically increasing in power with redshift so we may interperet a deficit with respect to the observed counts as a lack of evolutionary amplitude at lower redshifts while an excess with respect to the observed counts may be interpreted as an indication of too much evolution at higher redshifts.

At 15$\umu$m, the integral counts appear to be fitted well by all scenarios with the only discrepancy between the density and luminosity evolution models appearing at the faintest flux levels (remembering that evolution is also present in the other more established components of the cppRR model). Serjeant et al. \shortcite{serjeant00} found that the 15$\umu$m integral counts could indeed be well fitted by a variety of evolving models (cppRR, Franceschini et al.\shortcite{fran94}, Guideroni et al. \shortcite{guid97}, Xu et al.\shortcite{xu98}). However, examining the 15$\umu$m differential counts provides a more discriminating picture. The differential counts effectively measure the slope of the integral counts and are hence more sensitive to subtle changes that may not be prominent in the integral counts at first sight. Here, again we see a discrepancy between the two evolution scenarios with the density evolution model producing slightly higher differential counts (i.e. steeper integral counts) at the faintest fluxes (perhaps hinting at being marginally inconsistent with the observed counts due to too much power at higher redshifts?). However, what is more striking is that neither the classical density nor luminosity evolution models can reproduce the subtle changes of slope in the 15$\umu$m counts that are not easily apparent from the integral source counts alone. It is apparent from the differential counts that the counts steepen drastically from $\sim$3mJy-0.4mJy ($lg(dN/dS)= \alpha lgS, \alpha \approx -3$) above Euclidean expectations, turn over at $\sim$0.4mJy and rapidly flatten at fainter fluxes. The analysis by Serjeant et al. \shortcite{serjeant00} concluded that models incorporating pure luminosity evolution provided a better fit than the density evolving models although they assumed a lower value for the density evolution and moreover their observations did not extend down in detail below 10mJy in the differential counts.

At 60$\umu$m the IRAS data does not really extend to deep enough fluxes to gain meaningful insight from the models, with both the luminosity and density evolution scenarios fitting both the integral and differential counts, although the faintest differential counts would imply stronger evolution than currently assumed. 

However, at 170$\umu$m none of the models can account for the observed source counts at fainter fluxes (although the brightest points are fitted by all models). At 170$\umu$m we could envision a general increase in the magnitude of the evolution as generally fitting the observed counts (i.e. increasing the density evolution power law index $g$). However, in the case of density evolution, fitting the FIRBACK counts \cite{pug99} would result in violating the limits set by the fluctuation analysis in the Lockman Hole of Matsuhara et al.~\shortcite{mat00} if the trend of a flattening in the source counts is continued below fluxes of 70mJy. This again points to some constraint on the magnitude of the allowed density evolution if the evolution is allowed to increase monotonically to higher redshift. If density evolution is to account for the 170$\umu$m counts while simultaneously not violating the fluctuation analysis limits then it must take place a relatively low redshift (z$\sim$1). 

Finally, in the sub-mm at 850$\umu$m, the biggest deficit between the observed and predicted counts is seen. Surveys of both blank fields and around lensed clusters  ~\cite{hugh98}, ~\cite{smail97},~\cite{barg98}, ~\cite{eales99} of areas ranging from 0.002-0.12sq.deg., from 1-8mJy have revealed source densities up to 3 orders of magnitudes above no-evolution predictions. Furthermore due to the steep K-corrections in the sub-mm, the flux of high redshift galaxies is enhanced, resulting in almost no difference in the ability to detect a galaxy between redshift 1-10. However, even at 850$\umu$m we also see that monotonically increasing power law density evolution cannot fit the counts and produces too much evolution at higher redshift. Blain et al.~\shortcite{blain98} concurred with this result by finding that fitting the counts with  pure density evolution would result in a violation of the background radiation by 2 orders of magnitude. Interestingly however, is the luminosity evolution model that comes tentatively close to the observed source counts at the brightest and faintest fluxes.

From fig.~\ref{classics} it is apparent that significantly drastic evolution, in both magnitude and form, will be required to fit the observed source counts. From the fits to the faint ends of the 170$\umu$m, 850$\umu$m and marginally to the 15$\umu$m counts, significant density evolution at high redshifts seems unlikely. As well as exceeding the observed source counts at 170$\umu$m \& 850$\umu$m, such density evolution would also violate the constraints set by the FIR background radiation. Matsuhara et al. \shortcite{mat00} found that the integral counts at 170$\umu$m in the Lockman Hole at a flux of $\sim$150mJy had a steep slope, $lg(dN/dS)= \alpha lgS, \alpha \approx -3$. However they summized that if this slope did not become shallower before $\approx$1mJy then the power in the calculated fluctuation spectrum would exceed the observed power by a factor of 10. Therefore at some point the source counts must flatten to a value of $\alpha < -2$. Fig.~\ref{classics} shows that assuming monotonically increasing density evolution to z$\approx$ 2.5, the counts remain steep down to fluxes fainter than 1mJy. 
At 850$\umu$m, constraints set by COBE measurements of the sub-mm background radiation (Fixsen et al. \shortcite{fixsen98}, Hauser et al.\shortcite{hauser98}) require that the SCUBA sub-mm counts must converge at fluxes $\sim 0.3-0.5mJy$ \cite{hugh98}, \cite{hugh99}. Again, assuming power law density evolution, the counts remain steep down to fluxes deeper than 0.1mJy.  
Observations and follow ups of the ISO 15$\umu$m surveys of the CFRS, HDF, Lockman Hole \& Marano fields identified the sources with galaxies at redshifts of between 0.3 and 1 with a median redshift $\approx$0.8  \cite{flores99a}, \cite{mann97}, \cite{elbaz99}.

In summary,  there is too little density evolution at low redshift ($z\sim 1$), and too much density evolution at high redshift ($z\sim 2$) in the power law evolution parameterization to fit the observed source counts at far-IR and sub-mm wavelengths. Furthermore, the 15$\umu$m and 170$\umu$m counts require extreme evolution from a redshift of $\approx$0.2 (the limit of IRAS) to $\approx$1. On the other hand the SCUBA 850$\umu$m counts and the 170$\umu$m fluctuation analysis will constrain the upper redshift limit to the evolution. Therefore, I assume that the ISOCAM 15$\umu$m differential counts can provide a good boundary to constrain the evolution in the low redshift Universe ($0<z<1$) and the 850$\umu$m SCUBA sub-mm counts will constrain the evolution in the high redshift Universe ($1<z\sim 3$). The evolution of the galaxy population will also be constrained at high redshifts by the measurements of the cosmic infrared background (CIRB). By simultaneously examining evolution scenarios that successfully fit both the 850$\umu$m and 15$\umu$m counts it is expected that the model fits to the data at intermediate wavelengths and CIRB will also converge.

\subsection{Density Evolution Analysis}

\begin{figure}
\centering
\centerline{
\psfig{figure=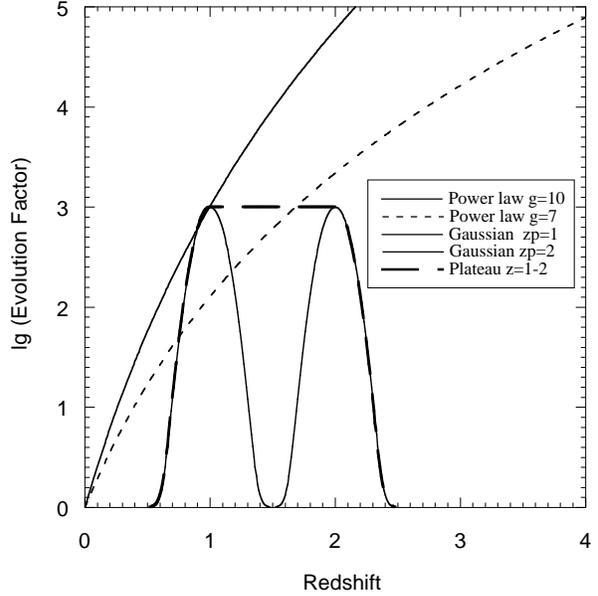,height=8cm}
}
\caption{Density evolution parameterizations for the ULIG component models discussed in the text. Density evolution is investigated via three exponential models. Two Gaussian parameterizations peaking at a redshift of 1 and 2 respectively plus an exponential plateau model ranging from z=1-2. Also shown for comparison are classical power law density evolution models with power indices g=10 and g=7 (i.e. that previously assumed for the classical density evolution of IRAS sources, see fig. 4) respectively. The magnitude of the density evolution in the zp=1 Gaussian scenario is equivalent to the magnitude of the evolution in the power law model using a value of g=10 at z=1.
\label{density}}
\end{figure}

Treating only the ULIG component, to investigate the density evolution I consider 3 simple Gaussian models (see Fig.~\ref{density}, although not exponential in the strictest sense, hereafter referred to as exponential evolution). Narrow Gaussian profiles peaking at a redshift of 1 and 2 are compared with each other and also with a scenario, referred to as the {\it Gaussian plateau}, similar to the {\it anvil} model of Blain et al.~\shortcite{blain99b}, that evolves exponentially to redshift $\sim$1 and declines exponentially after a redshift $\sim$2. The Gaussian profiles have the form,

\begin{equation}
   D(z) = 1 + g  exp[-{(z-z_{p})^{2}\over{2\sigma ^2}}],
\label{Dz}
\end{equation}

with the peak magnitude of the evolution, $g=1000$, the Gaussian width (effectively the evolution timescale) $\sigma =0.1$ and the peak redshift $z_{p}=1$ \& $2$ respectively. The Gaussian plateau model assumes an exponential increase and decrease as equation ~\ref{Dz}, to and from redshifts of 1 and 2 respectively and a constant value of $1+g$ in between. Although these models are merely intended to highlight the effect of density evolution at different redshifts, the narrow Gaussian profiles may be considered as epochs of short, intense merging while the Gaussian plateau represents a longer episode of galaxy merging (and consequent star formation / galaxy IR emission).

\begin{figure}
\centering
\centerline{
\psfig{figure=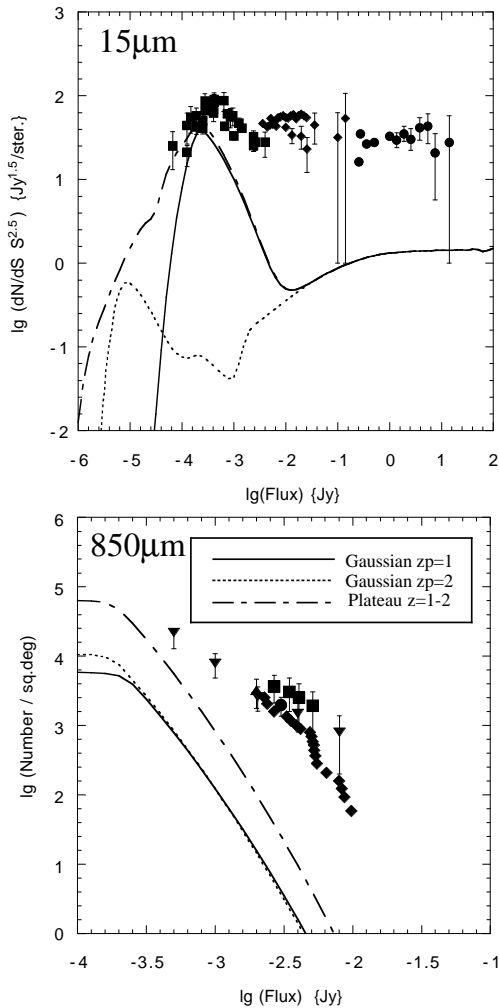,height=14cm}
}
\caption{Source counts of the high luminosity component only at 15$\umu$m (differential) and 850$\umu$m (integral) respectively. The 3 model lines correspond to the three exponential density evolution scenarios of the ULIG component discussed in the text. References for observed points are given in the text and figures.
\label{devol}}
\end{figure}

Using these density evolution scenarios, the corresponding evolving ULIG components are re-plotted in fig.~\ref{devol} at 15$\umu$m and 850$\umu$m. Several points are worth noting in comparison with fig.~\ref{classics}. Assuming the exponential evolution from equation~\ref{Dz}, a value of $g=1000$ at redshift 1 corresponds roughly to a $g \sim 10$ in the $(1+z)^{g}$ power law scenario with the significant difference that the high redshift density evolution has vanished. The 15$\umu$m counts remain relatively unchanged since they are not so sensitive to any density evolution at high redshift ($z>1$). However, the effect on the 850$\umu$m sub-mm counts is significant in that the excess at the faint fluxes ($\sim 1mJy$) has disappeared. In the case of the $z_{p}=2$ scenario, the 15$\umu$m counts cannot be fitted due to the fact that there is no evolution at lower redshift where it is needed. In fact, as can be seen from the Gaussian plateau model, the 15$\umu$m counts in general are clearly rather insensitive to density evolution at higher redshifts ($z > 1$). From the 15$\umu$m counts, it is clear that power in the local density evolution is required. For the 850$\umu$m counts (fig.~\ref{devol}), density evolution peaking at either z=1 or 2 is acceptable. However, the Gaussian plateau model, with constant density evolution from z=1-2 violates the faint end of the 850$\umu$m counts.

\subsection{Luminosity Evolution Analysis}

Despite the apparent success at 15$\umu$m, the brighter SCUBA counts at 850$\umu$m ($S>1mJy$) seem impossible to fit with this form of density evolution alone. Even the evolved counts still fall too far to the left and low in the $lg(N) / lg(S)$ plane when compared to the observations $> 1mJy$. In general, the brighter SCUBA population is comprised of fewer but more powerful sources (a high redshift 8mJy SCUBA source being approximately 10 more luminous than the most luminous local sources). Additional clues come from a further inspection of the luminosity evolution models of fig.~\ref{classics}. In general the addition of luminosity evolution has the generic effect of shifting the source counts towards brighter fluxes (to the right of the $lg(N) / lg(S)$ plane (as the brighter, evolved sources are seen at lower sensitivities i.e. higher flux values). 

Therefore, I extend the analysis of the density evolution scenarios to include a similar subset of luminosity evolution models. Following the same doctrine as the density evolution analysis I define the luminosity evolution as, 

\begin{equation}
   L(z) = 1+ k exp[-{(z-z_{p})^{2}\over{2\sigma ^2}}],
\label{Lz}
\end{equation}

with the magnitude of the evolution, $k=5$, $\sigma =0.3$, $z_{p}=1$ \& $2$ respectively. Thus this form would be equivalent to a value of $k \sim 2.3$ for the case of $(1+z)^{k}$ power law luminosity evolution. Note the disparity in the magnitude / width of the luminosity and density distributions. It should be noted that, firstly, the luminosity and density evolution enter the cosmological equations differently. Density evolution increases the number of sources observed in a given volume defined by a sensitivity $S(z)$ and redshift, $z$. The volume remains the same and the number density of sources increase. On the other hand the effect of luminosity evolution is to increase the observable volume via a lowering of $S(z)$ while preserving the source number density. Furthermore, the luminosity evolution analysis is performed on the assumption of there already being the presence of density evolution at low redshift manifested as the $z_{p}=1$ exponential density evolution scenario.

\begin{figure*}
\centering
\centerline{
\psfig{figure=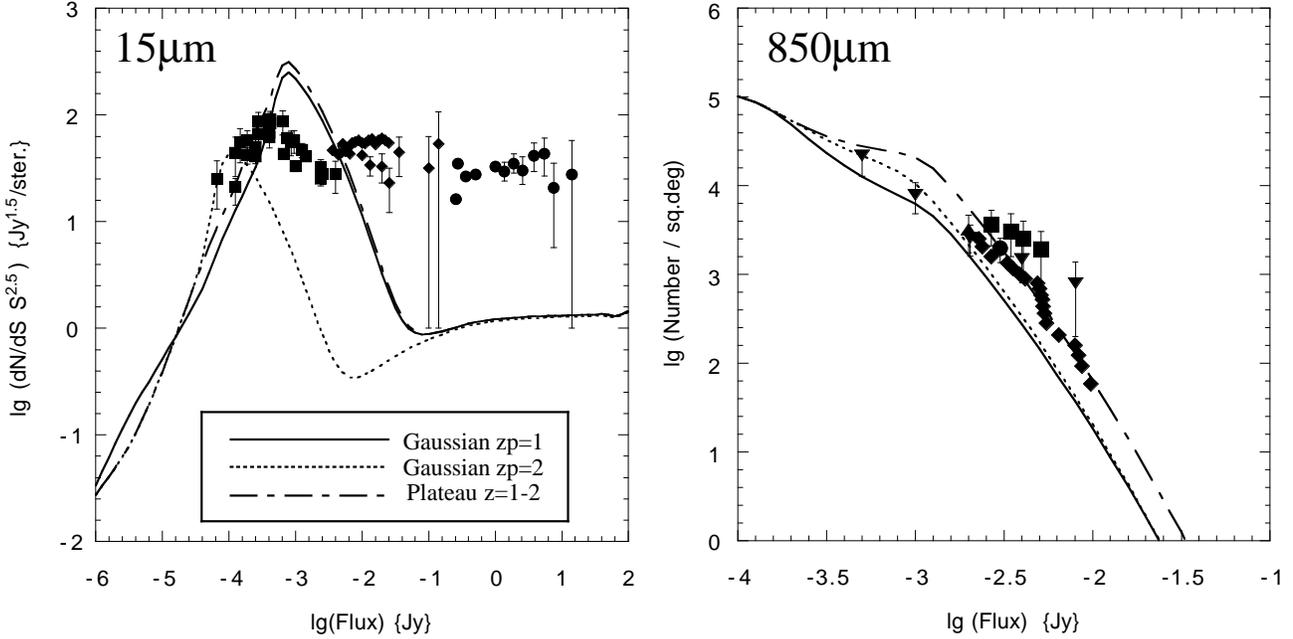,height=13cm}
}
\caption{Source counts of the high luminosity component only at 15$\umu$m (differential) and 850$\umu$m (integral) respectively. The 3 model lines correspond to the three exponential luminosity evolution scenarios discussed in the text. Note that the low redshift density evolution scenario is also assumed and included in these plots. References for observed points are given in the text and figures.
\label{levol}} 
\end{figure*}

Fig.~\ref{levol} shows the differential and integral counts of the ULIG component at 15$\umu$m and 850$\umu$m respectively for the 3 luminosity evolution scenarios described by eqn.~\ref{Lz}. In this analysis, density evolution of the form in eqn.~\ref{Dz} is assumed with parameter values of $g=500$, $\sigma =0.1$, $z_{p}=1$. As expected, the addition of luminosity evolution moves the counts to the right in the $lg(N) / lg(S)$ plane.

At 15$\umu$m, evolution at higher redshift ($z_{p}=2$) is not well constrained by the counts as the 15$\umu$m sources are expected in general to reside at lower redshift. The $z_{p}=1$ Gaussian and plateau (due to the z$\sim$1 contribution) scenarios, strongly violate the differential counts at 15$\umu$m not in fact due to the strength of the luminosity evolution alone but more due to the superposition of both the luminosity and density evolution effects at these redshifts. The 15$\umu$m differential counts put a severe constraint on the magnitude of the luminosity evolution at lower redshift in the presence of density evolution.

At 850$\umu$m, the effect of the addition of luminosity evolution is quite spectacular. The shift to the right in the $lg(N) / lg(S)$ plane can effectively fit the SCUBA counts at fluxes both fainter and brighter than 2mJy. There is little difference between the high or low redshift evolutionary scenarios implying that many of the SCUBA sources are at redshifts higher than unity allowing the $z_{p}=2$ scenario to compete with the combined effect of the luminosity and density evolution in the $z_{p}=1$ scenario. However, it should be noted that there is a significant difference between the two Gaussian scenarios and the plateau scenario. It would seem that constant luminosity evolution over the entire redshift range z=1-2 is ruled out by the count constraints. 

As with the density evolution analysis, the SCUBA 850$\umu$m counts are constrained by the evolution at higher redshift while the 15$\umu$m counts are severely constrained at lower redshift. This bipolar constraint is highlighted in the way that the plateau curve {\it hugs} the $z_{p}=1$ Gaussian luminosity evolution curve at 15$\umu$m, while at 850$\umu$m there is a more clearer superposition of the 2 Gaussian scenarios.

\section{Construction of New Evolutionary Models}\label{sec:counts}

\subsection{Source Counts}

From the analysis of the effects and constraints from density and luminosity evolution of the IR high luminosity population (assumed to be ULIGs) in the previous section, the following points / constraints can be made,

\begin{enumerate}
  \item The evolution can be constrained by simultaneous consideration of the 15$\umu$m and 850$\umu$m counts. \\
  \item The 15$\umu$m counts are predominantly dominated by low redshift galaxies and insensitive to the high redshift population. \\
  \item The faint end of the 850$\umu$m counts constrains the amount of density evolution allowed. \\
  \item Simple density evolution scenarios can fit the 15$\umu$m counts but are unable to account for the sub-mm counts at fluxes $>2mJy$. \\
  \item The addition of luminosity evolution has the effect of moving the source counts towards brighter fluxes in the $lg(N) / lg(S)$ plane. \\
  \item The 15$\umu$m counts place strong constraints on the luminosity evolution at low redshift but are insensitive to luminosity evolution at higher redshift . \\
  \item Luminosity evolution at either high or low redshift can fit the 850$\umu$m counts (i.e. the SCUBA counts effectively only constrain the magnitude of the luminosity evolution in this scenario). \\
\end{enumerate}

From these considerations a model can be envisioned that includes both density evolution at lower redshift (peaking at $z \sim 1$) and luminosity evolution peaking at higher redshift. The exact details of the evolution are constrained by fits to the source counts which will depend on both the magnitude of the evolution and the degree of {\it overlap} in the redshift space between the tails of the density and luminosity evolution. The magnitude of the evolution at high redshift will ultimately be constrained by the cosmic infrared background.

\begin{figure*}
\centering
\centerline{
\psfig{figure=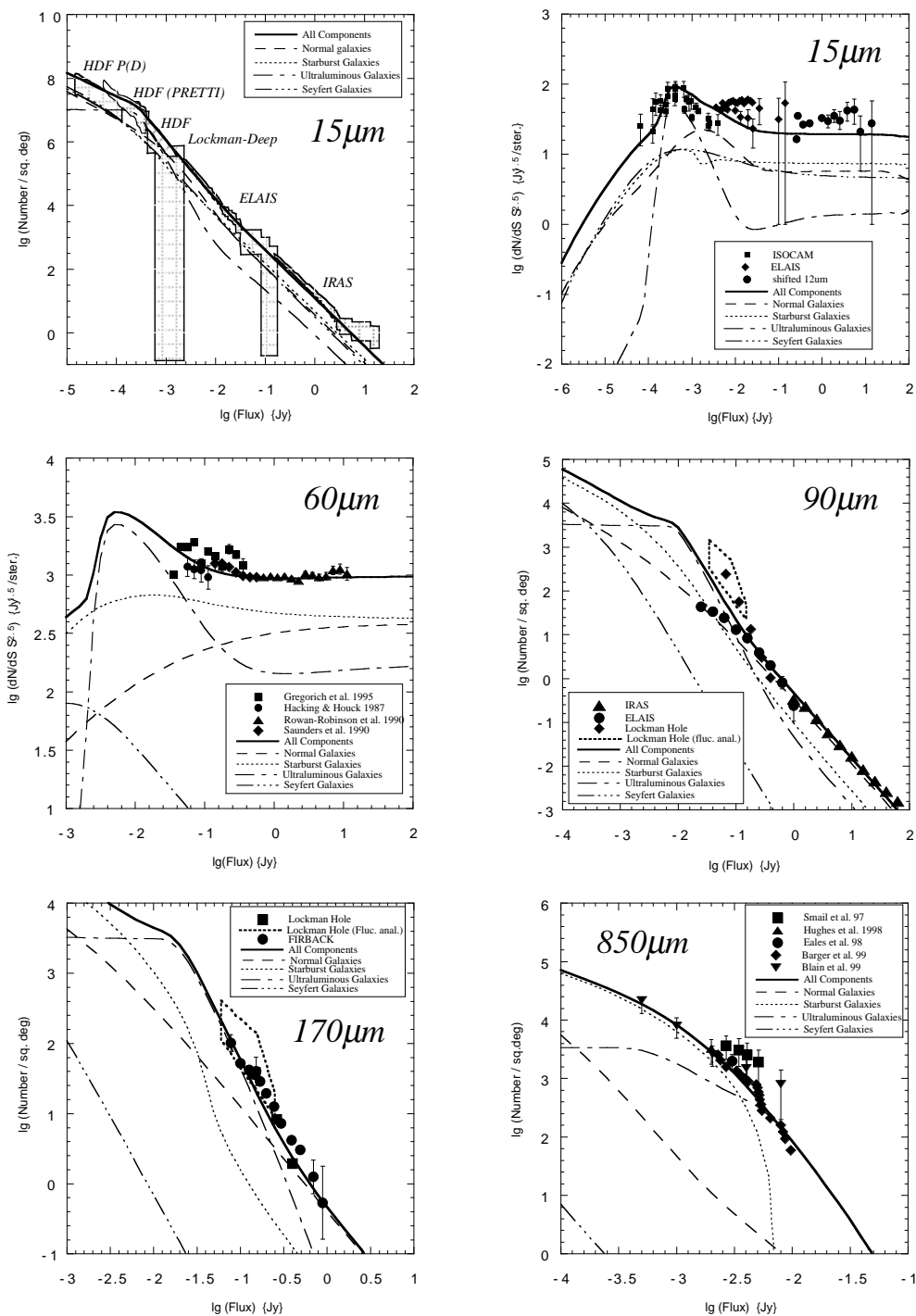,height=20cm}
}
\caption{Integral and differential counts at 15$\umu$m, 60$\umu$m, 90$\umu$m 170$\umu$m \& 850$\umu$m for the new evolutionary model of normal, starburst, ultraluminous and Seyfert galaxies as described in the text. Data is from 15$\umu$m - Oliver et al. (1997), Aussel et al. (1999), Serjeant et al. (2000), Elbaz et al. (1999). 60$\umu$m - Hacking \& Houck (1987), Rowan-Robinson et al. (1990), Saunders (1990), Gregorich et al. (1995). 90$\umu$m - Matsuhara et al. (2000) (fluctuation analysis - {\it dotted box\,}) , Efstathiou et al. (2000). 170$\umu$m - Kawara et al. (1998), Puget et al. (1999), Matsuhara et al. (2000) (fluctuation analysis - {\it dotted box\,}). 850$\umu$m - Smail et al. (1997), Hughes et al. (1998), Eales et al. (1999), Barger et al. (1999), Blain et al. (1999).
\label{newcount}} 
\end{figure*}  

Using these 3 constraints (15$\umu$m differential counts, 850$\umu$m integral counts and the CIRB) and the results from the previous analysis, the following evolutionary scenario is presented. The basic framework of the cppRR model is retained with the galaxy population consisting of 4 main components. Normal (cirrus) galaxies are represented by the {\it cool} 60$\umu$m IRAS luminosity function and are essentially non-evolving although a gradual decline towards higher redshift is included to represent gradual formation and/or transition through a starburst phase. The {\it warm} IRAS galaxies (starbursts) and Seyfert galaxies are represented by the 60$\umu$m and 12$\umu$m IRAS luminosity functions respectively and evolve in luminosity at a rate $\propto (1+z)^{3.2}$ to z=2.5 and then gradually decline at higher redshifts thus eliminating the artificial cut off employed previously in the cppRR model and pointed out by Rowan-Robinson~\shortcite{RR00} as being a general shortcoming with many models employing the {\it backward evolution} approach to galaxy evolution. Choosing a peak in the luminosity evolution at high redshift (z=2.5) is consistent with the higher redshift galaxies contributing more to the counts/background at longer wavelengths \cite{gispert00} . The high luminosity end of the IR 60$\umu$m luminosity function, the LIG/ULIG component, $lgL*_{60\mu m} \sim 11.6 L_{\sun}$, evolves in both number density and luminosity. The density evolution is strong and exponential, following eqn.~\ref{Dz} peaking at a redshift of 1 with the magnitude of the evolution, $g=220$ and Gaussian width, $\sigma =0.25$. The luminosity evolution rises exponentially as eqn.~\ref{Lz} with a magnitude $k=40$, $\sigma =0.58$ to a maximum redshift, $z_{p}=2.5$ and then slowly declines exponentially to higher redshift. The magnitude of the luminosity evolution at the peak redshift in the starburst, Seyfert and ULIG components is similar although the latter suffers a steeper decline towards lower redshift reflecting the more violent star formation within these galaxies. 

Fig.~\ref{newcount} shows the fits of the new evolutionary model to the source counts as observed by ISO (15, 90, 170$\umu$m), IRAS (60$\umu$m) \& SCUBA (850$\umu$m). The new model incorporating the strongly evolving LIG/ULIG component fits the observed counts extremely well with the LIG/ULIG component accounting for the bumps in the 15$\umu$m counts at 0.1-1mJy, the rise towards fainter fluxes in the 60$\umu$m differential counts, the excesses at 90$\umu$m \& 170$\umu$m and much of the SCUBA population brighter than 2mJy at 850$\umu$m. At 170$\umu$m Matsuhara et al. \shortcite{mat00} concluded on the basis of the 170$\umu$m/90$\umu$m colours of galaxies in their Lockman Hole field, that the faint sources could not be cirrus dominated (normal galaxies) but had to be starforming systems, consistent with the model prediction of a change in the dominant population, from cirrus to star forming ULIGs at $\sim$400mJy. Note that there is still significant discrepancy between the 90$\umu$m ISO source counts conducted by the the Japanese team (Matsuhara et al.~\shortcite{mat00}, Kawara et al.~\shortcite{kawara00}) and the ELAIS survey \cite{esf003}. The model predicts a value conservatively between these 2 estimations.

\subsection{Number Redshift Distributions}

\begin{figure}
\centering
\centerline{
\psfig{figure=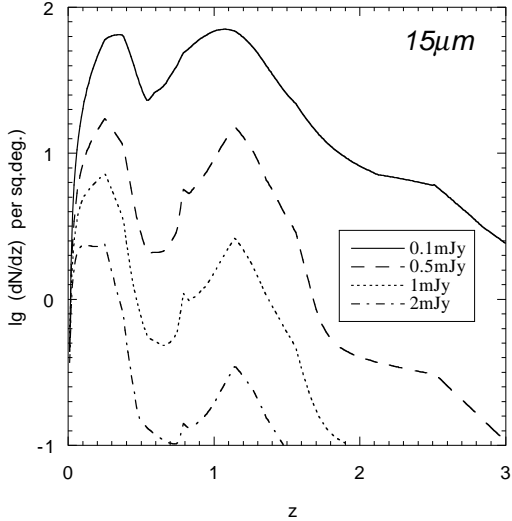,height=7cm}
}
\caption{Number-redshift distributions at 15$\umu$m for all components in the model for fluxes ranging from 0.1-2mJy. The peak at the lowest redshifts is due to the normal galaxy population. The ULIG population dominates at z$\sim$1.
\label{nz15}}
\end{figure}

\begin{figure}
\centering
\centerline{
\psfig{figure=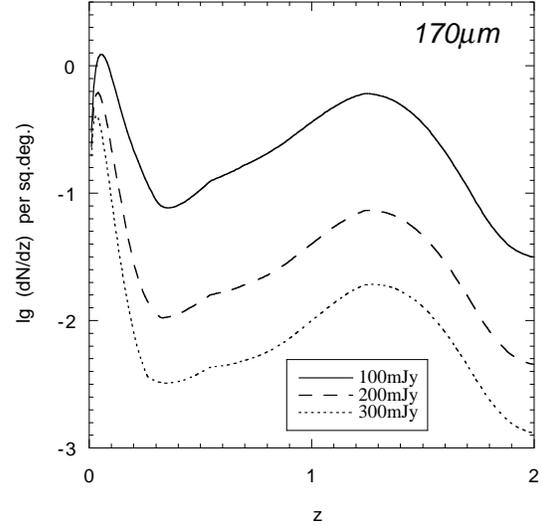,height=7cm}
}
\caption{Number-redshift distributions at 170$\umu$m for all components in the model for fluxes ranging from 100-300mJy. The peak at low redshifts $\sim$0.1 is due to normal galaxies. Note that the N-z distribution is measured as dN/dz, therefore although the peak due to normal galaxies is high, the width of its distribution is narrow.
\label{nz170}}
\end{figure}

\begin{figure*}
\centering
\centerline{
\psfig{figure=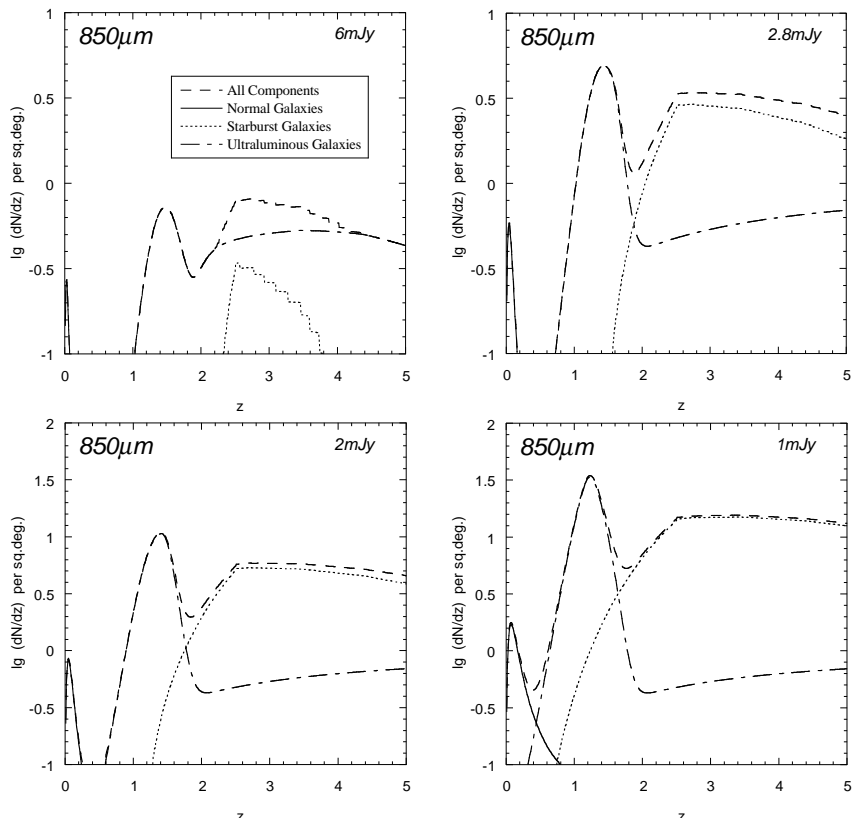,height=11cm}
}
\caption{Number-redshift distributions at 850$\umu$m for individual components in the model for fluxes ranging from 1-6mJy. In general, the normal galaxies are represented by the low redshift peak in the N-z distribution.
\label{nz850}} 
\end{figure*}

Figs.~\ref{nz15},~\ref{nz170},~\ref{nz850} show number redshift distributions for the new evolutionary model at 15, 170 \& 850$\umu$m respectively.

The numerous ISO surveys at 15$\umu$m have a general consensus with a strong evolution of sources to $z \sim 1$. The corresponding redshift distribution peaking in the range z$\sim$0.7-1.0 \cite{elbaz99b}.  The N-z distributions in figure~\ref{nz15} are broadly consistent with this strong evolution although due to the form of the evolution assumed there is a double peak in the distribution. The lower redshift peak is due to the normal galaxies that dominate the source counts to $\approx$1mJy. As the N-z distribution probes to fainter fluxes the contribution from the ULIG population (centred around z$\sim$1) increases. These models serve to show the effect of a strongly evolving population to z$\sim$1 on the source counts and moving the peak of the density evolution to lower redshift - or increasing the Gaussian width of the evolution would produce a smoother N-z distribution effectively merging the 2 peaks.

At 850$\umu$m, the redshift of the detected sources is notoriously difficult to measure unambiguously, the main problem being the 2-3\arcsec (or more) positional uncertainty in the sub-mm observations for a 15\arcsec resolution at 850$\umu$m ~\cite{hugh00}. Furthermore, there is often only a single definite detection at 850$\umu$m in the entire spectral range from the sub-mm to the far-IR $\sim$200$\umu$m . SCUBA preferentially selects galaxies at redshift $> 1$ due to the shape of the SED of galaxies in the sub-mm regime. A spectroscopic survey by Barger et al. ~\shortcite{barg99b} of a lensed sub-mm sample \cite{smail99} concluded that the sub-mm population resided mainly between z$\approx$1$\sim$3 with a median redshift between 1.5$\sim$2. For the 3 SCUBA galaxies for which there are CO spectroscopy confirmed redshifts, all lie at z$>$1 \cite{frayer98},\cite{frayer99},\cite{soucail99}.  However, the first identifications from the Canada-U.K. deep sub-mm survey have shown that the SEDs of the sub-mm sources (0.8, 15, 450, 850$\umu$m and 6cm) are consistent with high IR luminosity Arp 220 type luminous IR galaxies, many undergoing mergers/interactions with $\approx$ 1/3 of the sample at $z<1$, $1<z<3$ \& $z>3$ and a slightly higher median redshift of $\sim$2 \cite{eales99}, \cite{lilly99}. Eales et al~\shortcite{eales00} combined the Canada-UK deep survey, lensed surveys and HFF survey \cite{barg00}, producing a combined median redshift of 2.41. Finally Smail et al.~\shortcite{smail00} using the method of Carilli \& Yun ~\shortcite{carilli99},~\shortcite{carilli00} (see also Dunne, Clements \& Eales ~\shortcite{dunne00b}) combined radio data for a sub-mm sample to fit radio-sub-mm SEDs thus obtaining rough estimates of redshifts finding a median redshift between 2-3 for the sub-mm sources brighter than 1mJy. In general the trend is for a spread of redshifts between 1$\sim$3 and higher with a median around z$\geq$2. Figure ~\ref{nz850} is broadly consistent with these estimates. At the brightest fluxes ($\sim$6mJy) the redshift-distribution is dominated by the ULIG galaxy population consistent with the SCUBA observations of bright sources. In this region the evolution is dominated by the higher redshift luminosity evolution component. As the redshift distribution is sampled towards lower fluxes two points are worth noting. The first is the emergence of the starburst galaxy population as the dominant contributor at high redshift and the second is the strengthening of the number evolution component of the ULIG population.

\subsection{Background Radiation}

\begin{figure*}
\centering
\centerline{
\psfig{figure=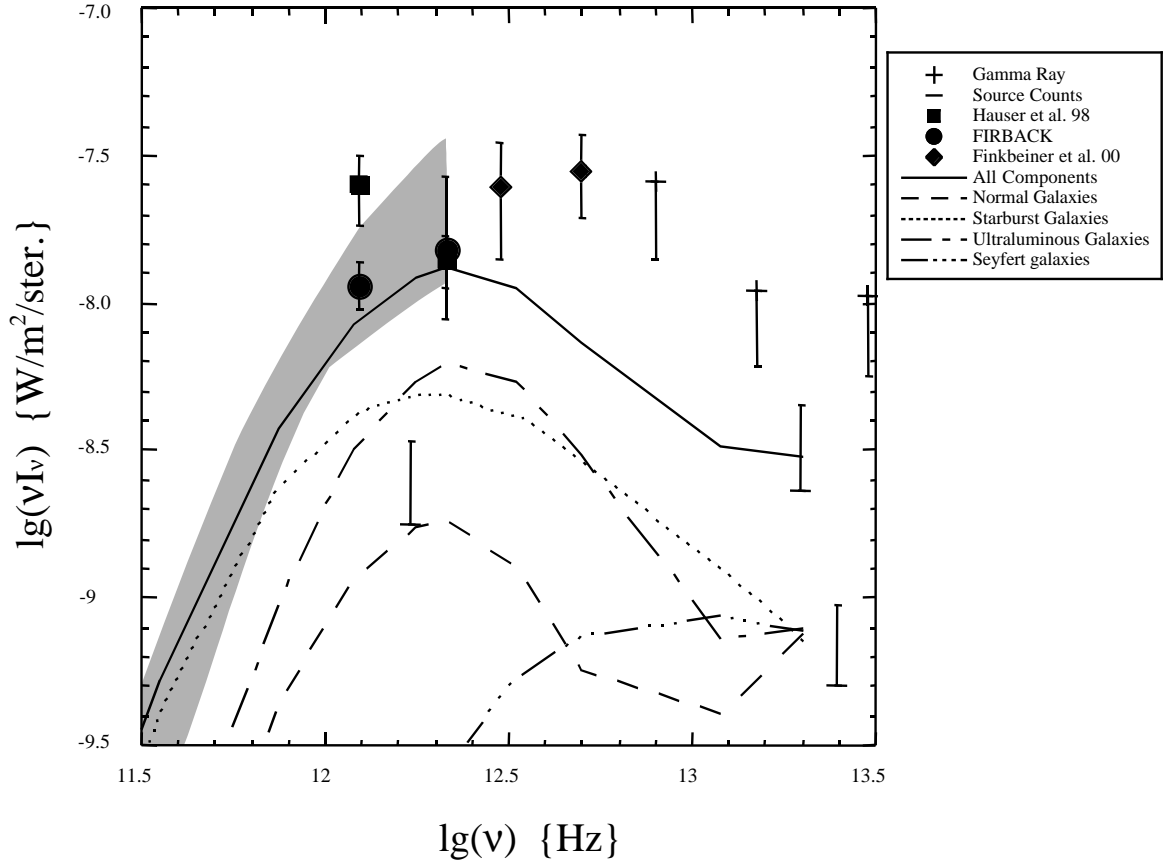,height=12cm}
}
\caption{The contributions to the infrared-sub-mm background light predicted by the model. The {\it Shaded area } in the figure represents the constraints set by the FIRAS experiment on COBE. {\it Source Counts } - refers to lower limits set from the integrated light of galaxy source counts. {\it Gamma Ray }  - refers to limits set by the attenuation of TeV gamma rays by the background radiation. {\it FIRBACK } - the integrated light of galaxies calculated from the FIRBACK survey (Puget et al. (1999)).
\label{cirb}} 
\end{figure*}

Further constraints on the evolutionary model come from the cosmic infrared background (CIRB) which manifests itself as the superposition of the unresolved population of faint sources in the Universe. The far-infrared background has been well constrained from $\sim 100\mu m - 2mm$ by measurements using the FIRAS/DIRBE instruments on COBE \cite{pug96}, \cite{fixsen98}, \cite{hauser98} and from ISO source count fluctuation analysis at 240-140$\umu$m \cite{pug99}, \cite{mat00}. Finkbeiner et al. ~\shortcite{fink00} have also reported possible detections and excess at 60 \& 100$\umu$m. In the sub-mm, lensed counts to 0.5mJy have discovered enough sources to approximately account for the 850$\umu$m background of $\nu I_{\nu}=5\times 10^{-1}nWm^{-2}sr^{-1}$ \cite{blain99}. In fig.~\ref{cirb}, the contribution of the new evolutionary model to the CIRB is shown compared with the observations described above. 

The model lies comfortably within the upper and lower constraints set by FIRAS. The cosmic background peaks at $\sim$140$\umu$m where detections by both Hauser et al.~\shortcite{hauser98} and the FIRBACK survey ~\cite{lagache99} are in good agreement. Note that the longer wavelength (240$\umu$m) measurements of COBE and FIRBACK do not agree so well although Lagache et al ~\shortcite{lagache99} assumed a significant correction due to warm interstellar dust (WIM) in their calculations. The model matches both the position and the magnitude of this peak extremely accurately. One direct consequence of this is that the majority of this peak emission would be expected to come from relatively low redshift galaxies \cite{gispert00}. The evolution in the high redshift regime is strongly constrained by the CIRB/Sub-mm background at longer wavelengths where the CIRB slope is shallower than the average galaxy SED implying a strong contribution from the integrated light of high redshift sources where the peak of the IR-SED is being sampled in the sub-mm waveband. It should be noted that assuming an $\Omega=1$ cosmology results in an extremely mariginal fit to the longer wavelength (sub-mm - mm) regime.

Constraints on the CIRB at shorter wavelengths come from DIRBE, IRTS ~\cite{mats00}, galaxy source counts and TeV gamma rays \cite{stecker00}, although the source count measurements and background measurements are yet to converge on an agreed value probably due to the difficulty in ascertaining the correction factor due to zodiacal light that peaks at $\sim$25$\umu$m \cite{ozernoy00}.

 Although the peak of the CIRB is well explained by the strongly evolving ULIG population in this model, the stronger contributor at sub-mm wavelengths are in fact the starburst galaxies. Furthermore although much of the sub-mm background may be made up from a population of ULIG galaxies (30-50$\%$ being resolved into individual galaxies with $S_{850}>2mJy$), the surface density of the sub-mm population is significantly higher \cite{hugh00}. Note that Eales et al.~\shortcite{eales00} have pointed out that the SCUBA fluxes may be upwardly biased by a factor of 1.4 due to source confusion and noise, resulting in a reduction of the resolved CIRB at 850$\umu$m to 20$\%$.

Such models of the CIRB are also consistent with the 850$\umu$m source counts (see fig.~\ref{newcount}). The ULIG component is completely dominant at bright fluxes down to $\sim$6mJy. Below this limit the starburst galaxies begin to contribute more strongly, where at $S_{850}<2mJy$ they become the dominant source population. Peacock et al.~\shortcite{peacock00} made a statistical analysis of the 850$\umu$m SCUBA map of the Hubble Deep Field (HDF) estimating that as much as 48$\%$ of the CIRB at 850$\umu$m could come from the sub-mm emission from the UV starforming galaxies with 30$\%$ from ULIG's \cite{hugh98} and a further 10$\%$ from AGN \cite{almaini99}. If the starburst population in the model is literally taken as the equivalent of the UV galaxies in the HDF then the model predictions slightly over predict this contribution from the starburst galaxies although the transition between starburst-LIG-ULIG is not clearly defined and therefore has a large scope for interpretation. The sub-mm source counts are also consistent with the results of Chapman et al.~\shortcite{chapman00} who carried out a sub-mm survey of 33 Lyman break galaxies \cite{madau96} to $\sim$1.3mJy. The average flux of these sources was found to be $S_{850} \sim 0.6mJy$ implying that they would occupy the fainter end of the sub-mm source counts and would not significantly contribute to the brighter $S_{850} \sim 5mJy$ source counts. Therefore it would seem that at the brighter sub-mm fluxes (2-5mJy) we are observing the ULIG population where as much as one third of the CIRB at 850$\umu$m can be resolved into individual bright sources. The underlying fainter population are distinct from the ULIGs and are more akin to the UV Lyman Break Galaxies and / or IRAS starburst galaxies. In this scenario the 850$\umu$m counts are a superposition of 2 galaxy populations, starbursts and ULIGs.

\section{Discussion}\label{sec:discuss}

The aim of this work has been to find an evolutionary model that can viably fit the galaxy source counts from the mid-IR to sub-mm wavelengths whilst not violating the constraints set by the IR background and CMB measurements and was not born from a fundemental physical assumption. 
The predicament of the model discussed in this paper is how does the bimodal evolution (i.e. dual evolution) manifest itself in the evolutionary history of a galaxy. Are we seeing a representation of two stages of galaxy evolution of the same population or different evolutionary paths of separate populations? Of the 19 SCUBA sources detected at 850$\umu$m in the Canada-UK sub-mm survey, only 2 were found to have ISO 15$\umu$m associations implying that the ISO and SCUBA sources are in fact different populations (or different redshift regimes) \cite{eales00}. Furthermore, what role do the ULIGs play in the evolutionary history of AGN? Do all QSOs go through a dusty ULIG phase seeing that the space density and luminosities of QSOs and ULIGs in the local Universe are quite similar ~\cite{soif87}. 

If this scenario is considered as 2 phases of evolution then the initial formation of the core (collapse of dark matter halo) would take place at high redshift accompanied by the initial formation of a black hole and associated star formation slowly declining towards lower redshift as the initial fuel is used up. The  peak in the co-moving number density of AGN is at redshifts $\sim$2, at lower redshifts these AGN are found to be hosted in giant elliptical galaxies ~\cite{mcclure99}. Thus the mid-IR 3-30um SED may be indicative of the galaxies  merging at lower redshifts activating / re-activating the central black hole where at some time the buried AGN expels the surrounding dust becoming the dominant energy source. The SED $>$50$\umu$m remains starburst dominated. The ULIGs would undergo rapid evolution from high redshift eventually evolving into today's local giant elliptical galaxies. Local giant ellipticals (gE galaxies) are the only local systems with core densities high enough ($100M_{\sun}pc^{-3}$) to account for a past ULIG evolutionary phase (Kormendy \& Sanders~\shortcite{korm92}, Binggeli~\shortcite{binggeli94}, Trentham~\shortcite{trentham00}).

\section{Conclusions}\label{sec:conc}

The {\it backward evolution} approach to galaxy evolution has been re-visited in light of the recent observations by ISO and SCUBA in the infra-red and sub-mm regions. Assuming classical evolutionary scenarios incorporating power law luminosity and density evolution it is impossible to simultaneously fit all the source counts from 15-850$\umu$m and the CIRB. 

A new evolutionary model is investigated which assumes the foundations of the Pearson \& Rowan-Robinson ~\shortcite{cpp96} model (cppRR). In addition to this original framework, a strongly evolving high luminosity IR galaxy component (referred to as the ULIG component but specifically the high luminosity end of the 60$\umu$m luminosity function) is added. This new population evolves in both density and luminosity and is capable of fitting both the source counts from 15-850$\umu$m and the cosmic far-infrared background radiation, including the peak at $\approx$140$\umu$m and the long wavelength tail.

In this scenario the ULIG component is tenuously interpreted as the progenitor of the present day giant elliptical galaxy population.

The excesses in the 15$\umu$m and 170$\umu$m counts are well explained by strong evolution to redshift $\sim$1 of a population of luminous, $lg L_{IR} \sim 12 L_{\sun}$, infrared galaxies. This population of galaxies becomes prominent at flux levels just below the limits of the IRAS 60$\umu$m survey (although previously hinted at) and can be associated with the archetypical ULIG Arp220. These results are generally consistent with the numerous surveys at 15$\umu$m ~\cite{elbaz00} which found that the dominant population above a redshift of $\sim$0.5 were massive ($M>10^{11}M_{\sun}$, luminous $L_{IR} > 10^{11-12}L_{\sun}$ galaxies.

At 850$\umu$m, these galaxies dominate the bright SCUBA counts at S$\geq$6mJy. Towards fainter fluxes the starburst galaxies begin to contribute strongly to the source counts, becoming dominant at fluxes $<2mJy$. In this scenario the sub-mm counts can be divided into 2 distinct populations, the dusty, giant ULIGs which would be considered the progenitors of giant elliptical galaxies and the IRAS starburst population / Lyman break galaxies.   
This conclusion is in good agreement with the analysis of Busswell \& Shanks~\shortcite{busswell00} who transformed the optical counts of faint starburst galaxies to the sub-mm waveband. Using a Bruzual \& Charlot \cite{bruzchar93} galaxy model with an exponentially decaying SFR of $\tau = 9Gyr$, their models successfully predicted optical results, including the Lyman break galaxies \cite{metcalfe00}. Furthermore they accounted for the sub-mm counts below a sensitivity of $\sim$1mJy but failed to fit the counts at brighter fluxes, concluding that an additional population was required (ULIGs/AGN?). Of further interest is that their models were also deficient in fitting the CIRB at 140-240$\umu$m where the ULIG population is most prominent.

Note, analysis of the so called {\it Madau} plot of the star formation history of the galaxies in this evolutionary model will be addressed in a later paper. The primary aim of this work is to ascertain what evolutionary scenarios can succesfully fit the observed source counts / background radiation.

From an analysis of luminosity and density evolution the following conclusions can be drawn from the galaxy source counts.

\begin{enumerate}
  \item The mid-IR region (15$\umu$m) is sensitive to evolution in the low redshift population but is not strongly constrained by evolution at higher redshift (z$>$1)
  \item The far-IR region at 60-90$\umu$m does not presently include enough information to constrain the evolution in the source counts.
  \item The far-IR region at 170$\umu$m is sensitive to the low redshift population and the faint end counts constrain the contribution to the peak of the CIRB.
  \item The sub-mm counts constrain the magnitude of the evolution in the high redshift Universe z$>$1 but due to the steep K-corrections, cannot distinguish evolutionary forms between redshifts 1-2. 
\end{enumerate}

Using the short wavelength 15$\umu$m and long wavelength 850$\umu$m counts, together with the CIRB as bounding conditions, the evolution over a wide range of both wavelength and redshift can be constrained. It is found that a scenario where the ULIG population undergoes 2 phases of evolution, a rapid merging epoch to z$\sim$1 and exponential evolution in luminosity to higher redshifts, best fits the observations.

In this scenario the high redshift ULIGs detected at 850$\umu$m fluxes $\geq$ 2mJy may be interpreted as the progenitors of todays giant elliptical galaxies while the population at fluxes fainter than $S_{850} \leq$2mJy represent the IRAS starburst population / Lyman break galaxies.  

Observations in the next few years in the mid-IR with SIRTF \cite{rieke00} and the Japanese ASTRO-F mission (Matsuhara et al.~\shortcite{mat98}, Watarai et al. ~\shortcite{wat00}, Pearson et al.~\shortcite{cpp01}), in the far-IR-sub-mm-mm with FIRST \cite{pilbrat00}, LMT \cite{schloerb97} and an all sky survey at far-IR wavelengths with ASTRO-F (Kawada et al.~\shortcite{kaw98}, Takahashi et al. \shortcite{takahashi00}) will undoubtebly solve many of these mysteries.

\section{Internet access to count models}

The source count models can be accessed through the world wide web at 

$http://www.ir.isas.ac.jp/\sim cpp/counts/$

\section{Acknowledgements}

CPP is supported by a Japan Society for the Promotion of Science (JSPS) fellowship. CPP thanks the referee Steve Eales and Hideo Matsuhara for significant comments and suggestions. CPP would like to thank Professor Haruyuki Okuda for providing him with the chance to spend a very fruitful and eventful time at the Institute of Space and Astronautical Science, Japan.



\bsp 

\label{lastpage}

\end{document}